\documentclass[referee]{raa}           
\usepackage{threeparttable}
\usepackage{graphicx,times}
\usepackage{natbib}
\usepackage{amssymb,amsmath}
\bibpunct{(}{)}{;}{a}{}{,}

\usepackage[pagebackref=true]{hyperref}

\begin{document}

\title{Expected evolution of the binary system ATLAS~J1138-5139}

 \volnopage{ {\bf 20XX} Vol.\ {\bf X} No. {\bf XX}, 000--000}
   \setcounter{page}{1}

   \author{Jing-Qi Chen
      \inst{1,2,3} ,
         Hai-Liang Chen \inst{1,3},
         Zheng-Wei Liu \inst{1,3},
         Xuefei Chen \inst{1,3},
         Zhanwen Han \inst{1,3}
   }

   \institute{ Yunnan Observatories, Chinese Academy of Sciences, Kunming 650216, China
   \\
   \and
   University of Chinese Academy of Sciences, Beijing 100049, China
             \\
        \and
        International Centre of Supernovae (ICESUN), Yunnan Key Laboratory of Supernova Research, Kunming 650216, P.~R.~China; {\it zwliu@ynao.ac.cn}
             \\
\vs \no
   {\small Received 20XX Month Day; accepted 20XX Month Day}
}

\abstract{
ATLAS~J1138-5139 is a newly detected ultra-compact double white dwarf (DWD) system which is composed of a $1.02\,M_{\odot}$ carbon-oxygen white dwarf (CO~WD) and a $0.24\,M_{\odot}$ helium (He) WD with an orbital period of about $27.68\,\mathrm{min}$, making it one of the shortest-period DWD systems known. The future evolution and final fate of this system remain unexplored. In this work, we investigate the evolution of ATLAS~J1138-5139 with the one-dimensional stellar evolution code \texttt{Modules for Experiments in Stellar Astrophysics (MESA)}. We find that ATLAS~J1138-5139 will evolve into an AM Canum Venaticorum (AM~CVn) system in about $\sim6.3\,\mathrm{Myr}$. Afterwards, the transferred material from the He~WD companion start to build up to form a He shell near the surface of the CO~WD. This accumulated He-shell masses can be up to approximately $0.12\,{M_{\odot}}$, which is likely to trigger a double-detonation (DDet) explosion of the CO~WD. We therefore expect that ATLAS~J1138-5139 will likely explode as a type Ia supernova eventually through the DDet explosion mechanism. Moreover, our calculations show that ATLAS~J1138-5139 will be a promising target for gravitational-wave (GW) detection by future detectors like \textit{LISA}, \textit{Tianqin} and \textit{Taiji}.
}

   \authorrunning{Chen, J.-Q., et al. }            
   \titlerunning{Evolution of ATLAS~J1138-5139}  
   \maketitle

%
\section{Introduction}           
\label{sect:intro}

Double white dwarfs (DWDs) are very important in many aspects of stellar physics. First, they have been widely used to place constraints on the common envelope evolution (CEE) and test the binary evolution theory \citep[e.g.][]{nvyp00,nt05,wivc12,2008AstL...34..620Y,2021ApJ...922..245B,ctch22}. Second, the DWDs with short orbital periods are potential targets for gravitational-wave (GW) detections by the GW detectors like \textit{Tianqin} \citep{Luo_2016},  \textit{Taiji} \citep{Ruan_2020} and the \textit{Laser Interferometer Space Antenna} (\textit{LISA}; \citealt{amaroseoane2017laserinterferometerspaceantenna,ebadi2024lisadoublewhitedwarf}). Third, the DWDs which are composed of a carbon-oxygen white dwarf (CO~WD) and a helium (He) WD could evolve to become the AM Canum Venaticorum (AM~CVn) stars if they experience stable mass transfer \citep[e.g.]{1986ApJ...311..753I,1995MNRAS.272..800H,1997ApJ...475..291I,2002ARep...46..667T,Nelemans_2004,2010PASP..122.1133S,2019ApJ...871..148L,Chen_2022}. Furthermore, if the He mass could steadily accumulate on the surface of the CO~WD during the evolution of an AM~CVn system, a He-shell detonation might be triggered, which will send a shockwave into the core of the CO~WD to trigger a second detonation to eventually cause a type Ia supernova (SN~Ia, plural SNe~Ia) through the double-detonation (DDet) explosion \citep[e.g.][]{1980SSRv...27..563N,1994ApJ...423..371W,bswn07,Guillochon_2010,Dan_2011,2014IAUS..298..269D,2014LRR....17....3P}. Fourth, the DWDs are also thought to be the potential progenitors of SNe~Ia via the merger channel \citep[e.g.][]{1990ApJ...348..647B,1995ApJ...438..887R,1997ApJ...481..355S,2004A&A...413..257G,2009A&A...500.1193L,2010ApJ...725..296F,2010Natur.463...61P,2012ApJ...747L..10P,2015ApJ...807..105S,Liu_2021,Liu_2022,Liu_2023}.

ATLAS~J1138-5139, which is also named as ``SMSS~J1138--5139'' by \citet{kosakowski2024newlisadetectabletypeia}, is a newly discovered nearby ultra-compact DWD that consists of a CO~WD and an extremely low-mass (ELM) He~WD companion with an orbital period of about 28 minutes at a distance of $D\sim553\,\rm{pc}$ \citep{chickles2024gravitationalwavedetectablecandidate,kosakowski2024newlisadetectabletypeia}. There are only six other ELM DWDs (up to orbital period of $30\,\mathrm{min}$) identified by spectroscopic surveys targeting \citep{Brown_2011,Brown_2020,Brown_2022,Kilic_2014,Kilic_2021}, and \citet{Burdge_2020} have identified 15 such binaries in a targeted search for ultra-compact binaries in the northern sky Zwicky Transient Facility (ZTF; \citep{Bellm_2019,Graham_2019,Masci_2019}) data. By performing a detailed analysis of photometry and spectroscopic observations of ATLAS~J1138-5139, \citet{chickles2024gravitationalwavedetectablecandidate} finds that ATLAS~J1138-5139 contains a $1.0\,M_{\odot}$ CO~WD and a low-mass He~WD companion ($\sim0.24\,M_{\odot}$), in which the CO~WD is acccreting mass from the low-mass He~WD. Their spectroscopic analysis confirms that the ongoing mass transfer is dominated by hydrogen (H), unequivocally pointing to a donor star that has not yet been stripped of its H-rich envelope. Once the H-rich layer is depleted, the donor star starts to transfer He-rich material to the CO~WD. They pointed that ATLAS~J1138-5139 would expect to either evolve into a stably mass-transferring AM~CVn system or trigger a low-luminosity SN~Ia within a few million years via the DDet explosion mechanism \citep{chickles2024gravitationalwavedetectablecandidate}. A parallel study by \citet{kosakowski2024newlisadetectabletypeia} supports their overall conclusion for ATLAS~J1138-5139, although their analysis reveals a slightly lower mass of the CO~WD of $0.99\,M_{\odot}$. Importantly, both studies highlight that the GW emission from ATLAS~J1138-5139 is expected to be detected by \textit{LISA} in the future due to its compact orbit. ATLAS~J1138-5139 might be the very first demonstration to blindly detect candidate SNe~Ia progenitor systems through GW signals alone. Despite the importance of ATLAS~J1138-5139 as a GW detectable candidate \citep{chickles2024gravitationalwavedetectablecandidate,kosakowski2024newlisadetectabletypeia}, its evolution and fate are still unclear, which requires a detailed binary evolution calculation based on its observed parameters.

The goal of this work is to investigate the future evolution and final fate of ATLAS~J1138-5139 by performing  one-dimensional (1D) detailed binary evolution calculation for long-term evolution of a binary system which has similar binary properties as ATLAS~J1138-5139. The article is structured as follows. In Section~\ref{sect:method}, we introduce the method and assumptions used in this work. In Section~\ref{sect:results}, we present the numerical results from our 1D detailed binary evolution calculation. We further predict the GW signals during the evolution of ATLAS~J1138-5139 and their detectability by future GW detectors. Finally, we summarize our results and conclusions in Section~\ref{sect:discussion}.

\section{METHOD AND ASSUMPTIONS}
\label{sect:method}

In this work, we investigate the evolution of ATLAS~J1138-5139 by using the stellar evolution code \texttt{Modules for Experiments in Stellar Astrophysics (MESA)}\citep{Paxton_2010,Paxton_2013,Paxton_2015,Paxton_2018,Paxton_2019}. We briefly describe the method and basic assumptions adopted in our \texttt{MESA} calculations in this section.

\subsection{Initial He~WD models}
\begin{figure}
\centering
\includegraphics[width=0.88\textwidth, angle=0]{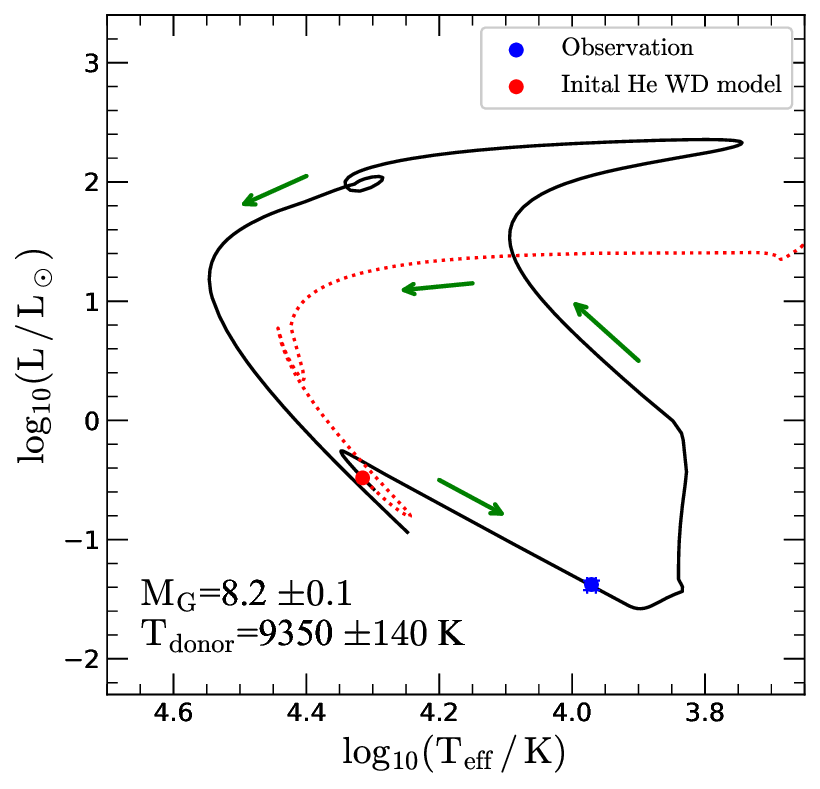}
\caption{Evolutionary track of our best initial He~WD model for  ATLAS~J1138-5139. The red dotted line represents the construction of a $0.255\,M_{\odot}$ He~WD which is composed of a $0.240\,M_{\odot}$ He core and a $0.015\,M_{\odot}$ H-rich envelope and the black solid line represents the subsequent evolution in the Hertzsprung–Russell (HR) diagram. The blue dot with error bars indicates the observed effective temperature and luminosity of the He~WD donor of ATLAS~J1138-5139. The red dot denotes the selected initial He-WD model for our binary evolution simulation.}
\label{Fig1}
\end{figure}

To investigate the evolution of ATLAS~J1138-5139, we construct an initial He~WD model based on the observed parameters given in Table~\ref{table2}. First, we construct a $1.2\,M_{\odot}$ zero-age main sequence (ZAMS) model with a solar metallicity of $Z=0.02$. Second, we evolve this ZAMS star until its He core mass reaches to a value of $0.24\,M_{\odot}$. At this point, we artificially remove the H envelope by the stellar wind with a constant wind mass-loss rate of $10^{-5}\,M_{\odot}\,{\rm yr}^{-1}$. The presence of H Balmer lines in spectra of ATLAS~J1138-5139 indicates that there is H-rich envelope onto the He~WD \citep{chickles2024gravitationalwavedetectablecandidate}. We therefore create a series of He~WD models with a range of H-shell masses from $0.01\,M_{\odot}$ to $0.05\,M_{\odot}$ for a given He-core mass of $0.24\,M_{\odot}$. Third, we evolve these He~WD models with \texttt{MESA} code as single stars. Fourth, we compare the evolutionary tracks of these single-star models with the observed luminosity and effective temperature of ATLAS~J1138-5139 to find the best matched model. Figure~\ref{Fig1} shows that a He~WD model with a residual H-envelope of $0.015\,M_{\odot}$ provides the best match to the observed effective temperature and luminosity of ATLAS~J1138-5139. We have tested models with H-envelope mass in the range of $0.01\,M_{\odot}-0.05\,M_{\odot}$. Our calculations suggest that variations in the H-envelope within this range do not lead to substantial differences in the subsequent evolution or the final outcome. We therefore select this model as the initial He~WD model for the subsequent investigation of the evolution of ATLAS~J1138-5139.

\begin{table}[ht]
\begin{threeparttable}
\centering
\caption[]{The observationally derived parameters of ATLAS~J1138-5139 given by \citet{chickles2024gravitationalwavedetectablecandidate} and \citet{kosakowski2024newlisadetectabletypeia}}
\begin{tabular}{lll}
  \hline\noalign{\smallskip}
Physical parameter &\citet{chickles2024gravitationalwavedetectablecandidate}
& \citet{kosakowski2024newlisadetectabletypeia} \\
\hline\noalign{\smallskip}
Orbital period ($P_\mathrm{orb}$) & $\mathrm{ 1660.92028(33)~s}$  & $\mathrm{27.6797 \pm 0.0009~min}$ (TESS)  \\ & $\sim$ 27.68~min & $\mathrm{27.69 \pm 0.03~min}$ (RV)  \\
\hline\noalign{\smallskip}
Orbital inclination ($i$) & $>76^\circ$ (Eclipses) & $\mathrm{88.7 \pm 0.1~deg}$ \\
 &  $\sim 88.6^\circ$ (Lightcurve model) &  \\
\hline\noalign{\smallskip}
Radial velocity of donor ($K_{\text{Donor}}$) & $\mathrm{687.4 \pm 3.8~km~s^{-1}}$  & $\mathrm{687 \pm 13~km~s^{-1}}$ \\
\hline\noalign{\smallskip}
Projected rotational velocity of donor ($\nu_{\text{min}}^{\text{Donor}}$) & $\mathrm{237.3 \pm 12.5~km~s^{-1}}$ \\
\hline\noalign{\smallskip}
Semi major axis ($a$) & $\mathrm{0.3262 \pm 0.0059\,R_\odot}$ \\
\hline\noalign{\smallskip}
Accretor mass ($M_{\text{WD}}$) & $\mathrm{1.02 \pm 0.04\,M_\odot}$ & $\mathrm{0.99 \pm 0.01\,M_\odot}$ \\
\hline\noalign{\smallskip}
Donor mass ($M_{\text{Donor}}$) & $\mathrm{0.24 \pm 0.03\,M_\odot}$ & $\mathrm{0.24 \pm 0.01\,M_\odot}$\\
\hline\noalign{\smallskip}
Donor radius ($R_{\text{Donor}}$) & $\mathrm{0.086 \pm 0.003\,R_\odot}$ & $\mathrm{0.0859 \pm 0.0005\,R_\odot}$ \\
\hline\noalign{\smallskip}
Donor temperature ($T_{\text{Donor}}$) & $\mathrm{9350 \pm 140~K}$ & $\mathrm{9650 \pm 300~K}$ \\
\noalign{\smallskip}\hline
\label{table2}  
\end{tabular}
\end{threeparttable}
\end{table}

\subsection{Binary star models and basic assumptions}

We utilize the test suite of \textit{star\_plus\_point\_mass} within the  \texttt{MESA} code for our binary evolution calculations. In our simulations, we do not resolve the detailed structures the CO~WD primary star and simply treat it as a point mass. We set the initial CO~WD mass and orbital period to be $0.99\,M_{\odot}$ and $0.02\,\mathrm{days}$, respectively. The observations of ATLAS~J1138-5139 indicate that it is undergoing the mass transfer \citep{chickles2024gravitationalwavedetectablecandidate,kosakowski2024newlisadetectabletypeia}. In our binary evolution model, we adopt an initial orbital period that is slightly longer than the observed value of ATLAS~J1138-5139. At the initial stage of the calculation, our initial model has not started the mass transfer yet. This setup allows our model to evolve for $10^{4}\;{\rm yrs}$ to naturally enter the mass transfer state that is consistent with the observed configuration of ATLAS~J1138-5139. As the system evolves, the He~WD starts to transfer mass to the CO~WD, shortening its orbital period to match the observed period and other properties of ATLAS~J1138-5139. Some basic assumptions in our binary evolution calculations are presented as follows.

\begin{enumerate}
\item[(1)]
We adopt the scheme of \cite{1988A&A...202...93R} to calculate the mass transfer rate in our binary evolution calculation:
\begin{equation}
    \centering
    \dot{M}_{\rm tr} \propto \frac{R_{\mathrm{RL}}^3}{M_\mathrm{d}} \exp \left( \frac{R_\mathrm{d} - R_{\mathrm{RL}}}{H_\mathrm{p}} \right),
\end{equation}
where $M_{\rm d}$ is the donor mass, $R_{\rm RL}$ is the Roche-lobe radius of the donor star, $R_{\rm d}$ is the radius of the donor star, and $H_{\rm p}$ is the pressure scale height.

\item[(2)]
We consider the angular momentum loss due to the GW radiation and the mass loss. The angular momentum loss due to the GW radiation is computed using the standard quadrupole approximation:
\begin{equation}\label{eq1}
  \frac{\mathrm{d} J_{\mathrm{gw}}}{\mathrm{~d} t}=-\frac{32}{5} \frac{G^{7 / 2}}{c^5} \frac{M_{\mathrm{a}}^2 M_{\mathrm{d}}^2\left(M_{\mathrm{a}}+M_{\mathrm{d}}\right)^{1 / 2}}{a^{7 / 2}},
\end{equation}
where $G$ is the gravitational constant, $c$ is the speed of light in vacuum, $a$ is the binary separation; $M_{\rm a}$ and $M_{\rm d}$ denote the masses of the accretor and donor, respectively.

\item[(3)]
 Our binary model experiences a H-rich mass transfer phase at the beginning due to a H-rich envelope onto the He~WD. We adopt the assumption that the mass transfer is conservative during this phase.

\item[(4)]

As the H-rich envelope of the He~WD is removed through mass transfer, it starts to transfer He-rich material onto the CO~WD. In this phase, we adopt the model given by \cite{Piersanti_2014} to calculate the accumulation efficiency of accreted He-rich material ($\eta_{\mathrm{He}}$) onto the CO~WD \citep[see also][]{1982ApJ...253..798N,Kato_2004}, which is given as follows.

\begin{equation}\label{eq3}
\eta_{\mathrm{He}}= \begin{cases}\dot{M}_{\mathrm{up}} / \dot{M}_{\mathrm{tr}} & \dot{M}_{\mathrm{tr}} \geq \dot{M}_{\mathrm{up}}, \\ 1 & \dot{M}_{\mathrm{up}}>\dot{M}_{\mathrm{tr}} \geq \dot{M}_{\mathrm{st}}, \\ 0 & \dot{M}_{\mathrm{st}}>\dot{M}_{\mathrm{tr}} \geq \dot{M}_{\mathrm{low}}, \\ 1 \ \  \mathrm{(no\ He\ burning)}& \dot{M}_{\mathrm{tr}}<\dot{M}_{\mathrm{low}}. \end{cases}
\end{equation}

Here, we follow the model of \cite{Piersanti_2014} and set $\dot{M}_{\mathrm{up}}=3.16 \times 10^{-6}\,M_{\odot}\,\rm{yr}^{-1}$, $\dot{M}_{\mathrm{st}}=5.89 \times 10^{-7}\,M_{\odot}\,\rm{yr}^{-1}$ and $\dot{M}_{\mathrm{low}}=4.90 \times 10^{-8}\,M_{\odot}\,\rm{yr}^{-1}$, respectively. First, if the mass transfer rate $(\dot{M}_{\rm tr})$ is larger than the critical mass transfer rate of $\dot{M}_{\rm up} = 3.16 \times 10^{-6}\;M_{\odot}\,{\rm yr}^{-1}$, we assume that the accreted He burns into CO stably onto the CO~WD at a rate of $\dot{M}_{\rm up}$. The unprocessed mass is lost in optically thick wind. Second, if the mass transfer rate is $\dot{M}_{\rm st} \leq \dot{M}_{\rm tr} < \dot{M}_{\rm up}$, either stable He burning or mild He flashes happen onto the CO~WD and we assume that there is no mass loss. Third, if $\dot{M}_{\rm low} \leq \dot{M}_{\rm tr} < \dot{M}_{\rm st}$, He-shell flashes are so strong that no mass can be accumulated by the CO~WD, i.e., the accumulation efficiency is zero. Fourth, if the mass transfer rate is lower than $\dot{M}_{\rm low} = 4.90 \times 10^{-8}\,M_{\odot}\,\mathrm{yr}^{-1}$, there is no He-burning and a He-shell is believed to build up near the surface of the CO~WD.

\item[(5)]
We simply assume that the CO~WD explodes as an SN~Ia through the DDet mechanism once the accumulated He-shell masses onto its surface reaches a critical value of $\sim 0.1\,M_{\odot}$ \citep{2010A&A...514A..53F,2011ApJ...734...38W,Shen_2018,Zenati_2023,Gronow2021}.

\end{enumerate}

\section{The results and discussions}
\label{sect:results}

\subsection{Binary evolution calculation}
\begin{figure}
    \centering
    \includegraphics[width=0.48\linewidth]{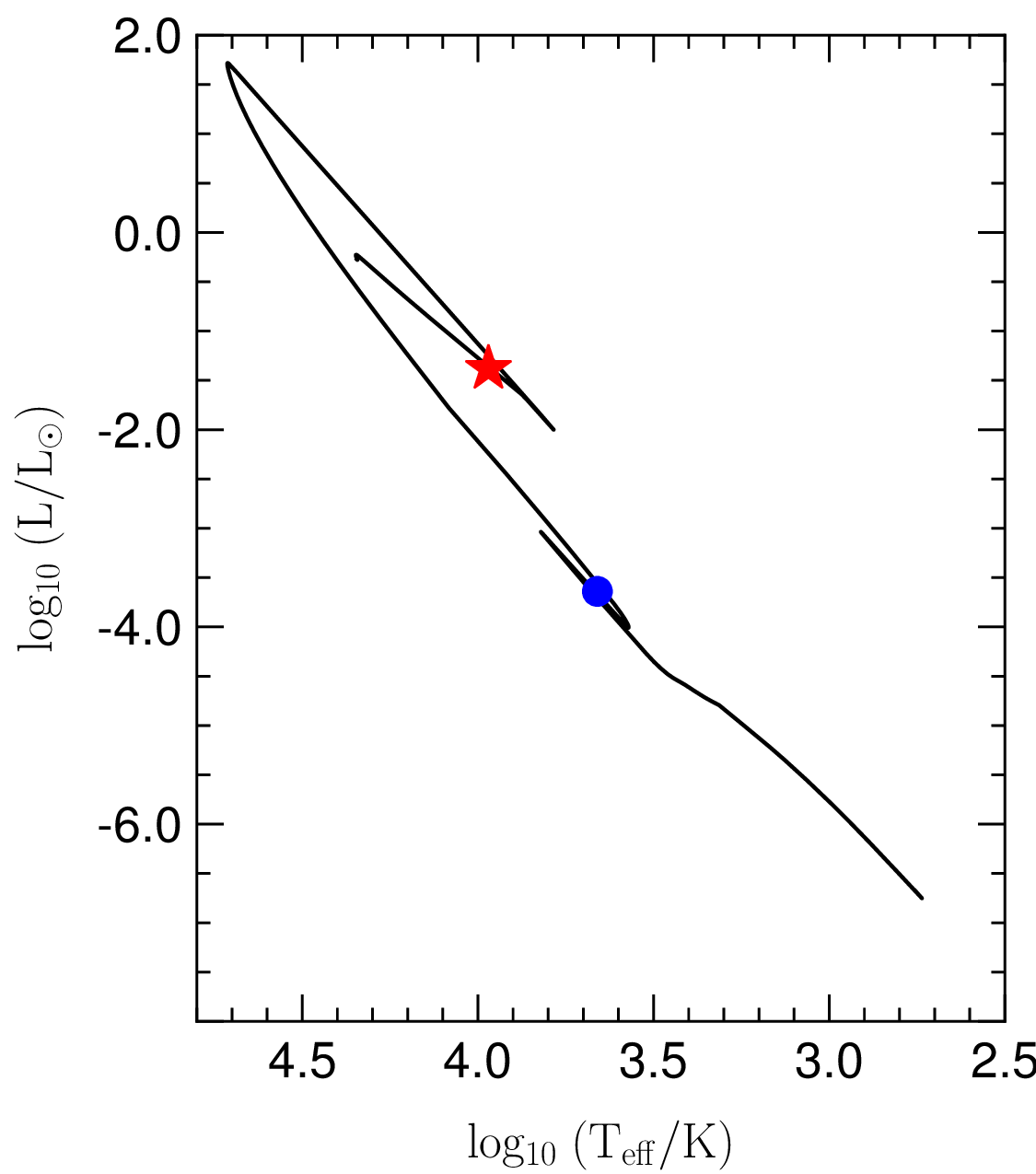}
    	\hfill
        \includegraphics[width=0.48\linewidth]{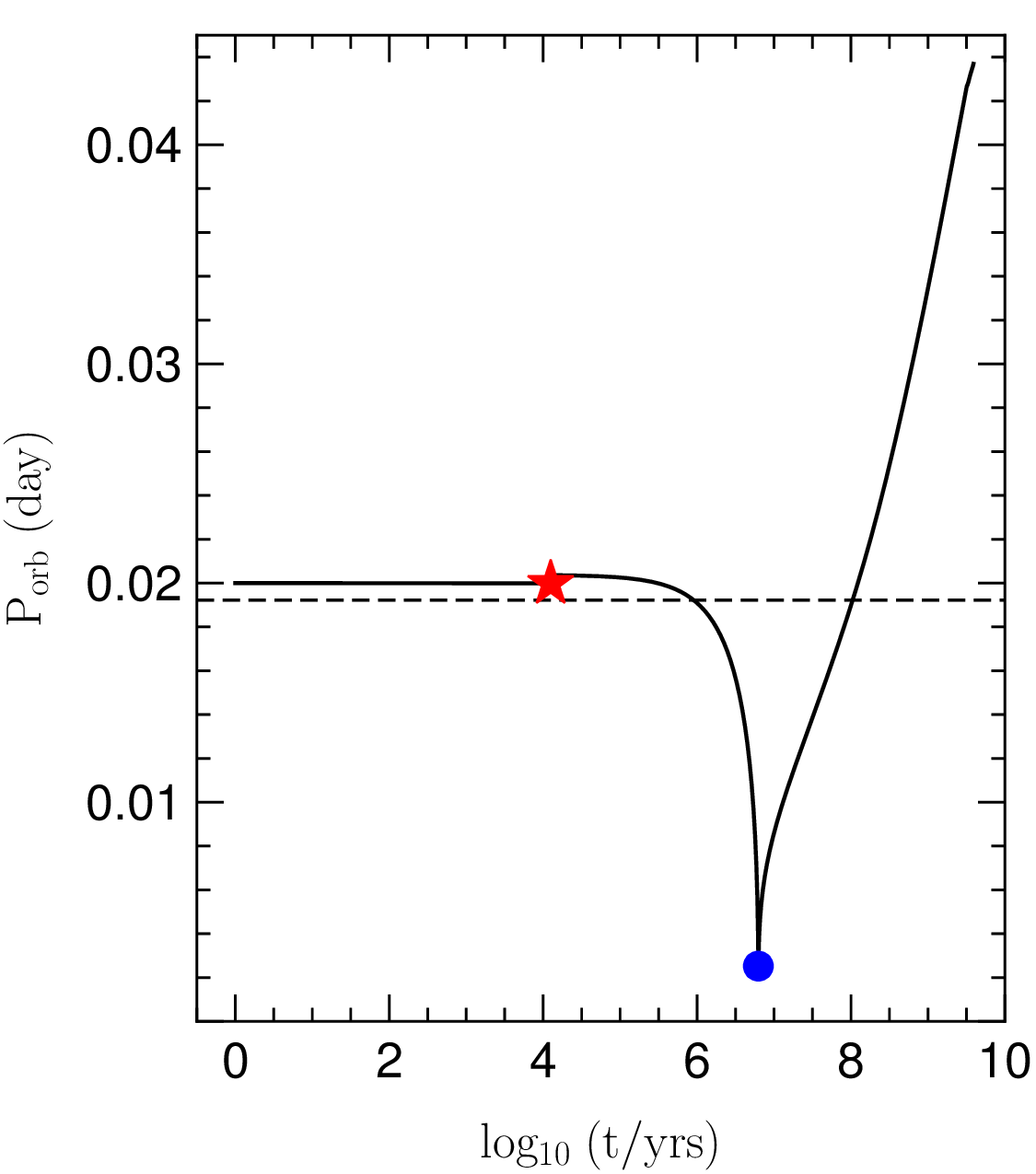}
    \caption{\textit{Left panel:} the evolution of the He~WD in HR diagram from our binary evolution calculation. The black line represents the evolution of temperature and luminosity of ATLAS~J1138-5139. \textit{Right panel:} the orbital periods of our model change over time. The black line represents the observed period of ATLAS~J1138-5139. Here, the red stars and blue dots respectively represent the observed properties of the donor star in ATLAS~J1138-5139 and the onset of He mass transfer of our model.}
    \label{fig:Evolution}
\end{figure}

In Figure~\ref{fig:Evolution}, we show the evolution of our binary star model in the HR diagram (left-hand panel) and orbital period-age diagram (right-hand panel), respectively. The spectroscopic observations of ATLAS~J1138-5139 indicate that the system is currently undergoing a H-dominated mass transfer \citep{chickles2024gravitationalwavedetectablecandidate,kosakowski2024newlisadetectabletypeia}. However, we set the initial orbital period of our model to be slightly longer than that of ATLAS~J1138-5139. This setup allows our model to consistently evolve for $t \sim 10^{4}\;{\rm yrs}$, at which the He~WD star expands to fill its Roche lobe to transfer H-rich material of its envelope to its companion star. At this stage, our model presents properties that are consistent with the current observed properties of ATLAS~J1138-5139. Subsequently, the He~WD begins to cool because that the mass of the H-rich envelope is too insufficient to sustain stable nuclear burning. During this phase, the orbital period of the binary system decreases due to the angular momentum loss through the GW radiation. As the orbital separation of this system continues to decrease, the He~WD fills its Roche lobe again at $t \sim 6.3 \times 10^{6}\;{\rm yrs}$. At this stage, the system starts the second mass transfer phase which is characterized by the He-dominated accretion. As a result, the system evolves to become an AM~CVn object. As the mass transfer continues, the mass of the He~WD decreases and its radius increases due to the adiabatic expansion in response to mass loss, which drives the widening of the binary orbit.

\begin{figure}
    \centering
    \includegraphics[width=0.88\linewidth]{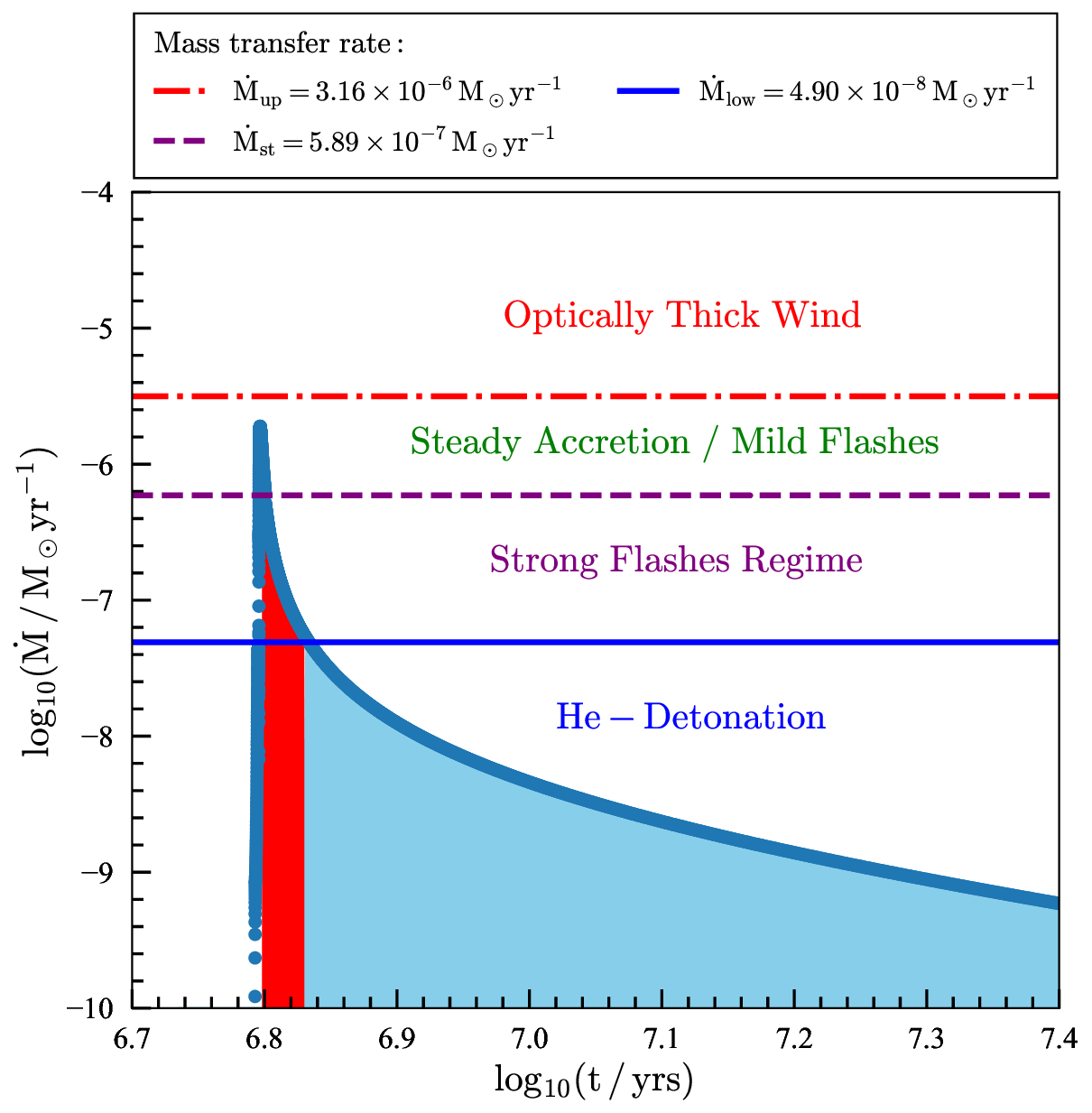}
        \caption{Mass transfer rate as a function of time in our binary evolution calculations. The blue dots represent temporal mass transfer rate. The horizontal lines indicate the critical rates for different accretion regimes given by \citet{Piersanti_2014}, which have been marked in the figure. The time interval from approximately $5\,\mathrm{Myr}$ to $11\,\mathrm{Myr}$ corresponds to the regimes in which He can't be accumulated because it is either stably burned or lost due to strong He-shell flashes (red shaded region). The subsequent interval from about $11\,\mathrm{Myr}$ to $25\,\mathrm{Myr}$ marks the phase in which the mass transfer rate falls below the He-burning threshold, allowing a He layer to build up on the surface of the CO~WD (blue shaded region).}
    \label{fig:mass_transfer}
\end{figure}

\begin{figure}
    \centering
    \includegraphics[width=0.8\linewidth]{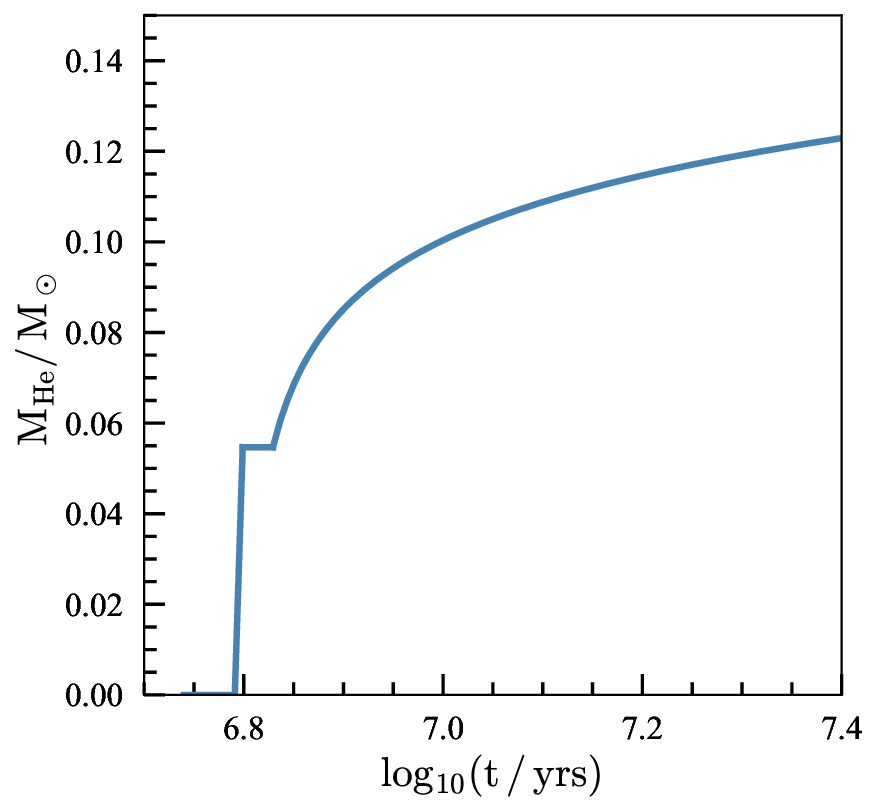}
    \caption{Total mass of the He-shell accumulated on the CO~WD as a function of time. In this work, we simply assume that the condition for the DDet explosion is satisfied once the accumulated He-shell mass onto the CO~WD exceeds $0.1\,M_{\odot}$, which means that an SN~Ia explosion occurs.
}
    \label{fig:he_cum}
\end{figure}

Figure~\ref{fig:mass_transfer} presents the evolution of mass transfer rate as a function of time after the H-rich envelope onto the He~WD is entirely stripped. As it is shown, the binary system enters different He-accretion regimes given by the model of \cite{Piersanti_2014}. When the mass transfer rate falls below $\dot{M}_{\rm low}=4.90 \times 10^{-8}\,M_{\odot}\,\rm{yr}^{-1}$, the transferred He-rich material stably accumulates on the surface of the CO~WD, i.e., no He-burning happens. In this work, we simply assume that a He-shell detonation would be triggered once the He-shell mass onto the CO~WD reaches a critical mass of $\sim0.1\,M_{\odot}$ \citep[e.g.,][]{Fink2007, Fink2010, Sim2010,2011ApJ...734...38W,Shen2018,Shen2021,Townsley2019,Boos2021, Gronow2021, Collins2022, Wong2023,Zenati_2023}, the resulting shock will compress the CO core to ignite the second detonation, leading to an SN~Ia explosion. Figure~\ref{fig:he_cum} shows the mass of a He-shell built up near the surface of the CO~WD changes over time. In our simulations, the accumulated He-shell mass can reach up to about $M_{\mathrm He}=0.12\,M_{\odot}$. We therefore predict that ATLAS~J1138-5139 will explode as an SN~Ia though the DDet explosion mechanism at the end of its evolution.

\subsection{Detectability of gravitational wave signal}

DWDs are important targets for future space-based GW detectors like \textit{TianQin} and \textit{LISA}. Additionally, the proposed \textit{Taiji} mission is also expected to be sensitive to such mHz GW sources, further enhancing the detectability of systems like ATLAS~J1138-5139. ATLAS~J1138-5139 has an extremely compact orbit and the ongoing mass transfer, making it a verification source that offers an unique opportunity to link the GW detection to the progenitors of SNe~Ia. To give the GW signal expected from ATLAS~J1138-5139, we calculate its chirp mass as follows:

\begin{equation}\label{eq4}
M_{\mathrm{chirp}}=\frac{\left(M_{\mathrm{CO}} M_{\mathrm{He}}\right)^{3 / 5}}{\left(M_{\mathrm{CO}}+M_{\mathrm{He}}\right)^{1 / 5}},
\end{equation}
where $M_{\mathrm{CO}}$ and $M_{\mathrm{He}}$ are the masses of the CO~WD and the He~WD, respectively.

Following the description given by \cite{Chen_2020}, we further calculate the characteristic strain of GW signal as follows:

\begin{equation}\label{eq5}
h_c \approx 2.5 \times 10^{-20}\left(\frac{f_{\mathrm{GW}}}{1 \mathrm{mHz}}\right)^{7 / 6}\left(\frac{M_{\text {chirp }}}{1 M_{\odot}}\right)^{5 / 3}\left(\frac{15 \mathrm{kpc}}{d}\right),
\end{equation}
where $d$ is the distance between the observed binary system and a GW detector, and the GW frequency of the binary system is defined as $f_{\mathrm{GW}}=2 / P_\mathrm{orb}$.

\begin{figure}
    \centering
    \includegraphics[width=0.95\linewidth]{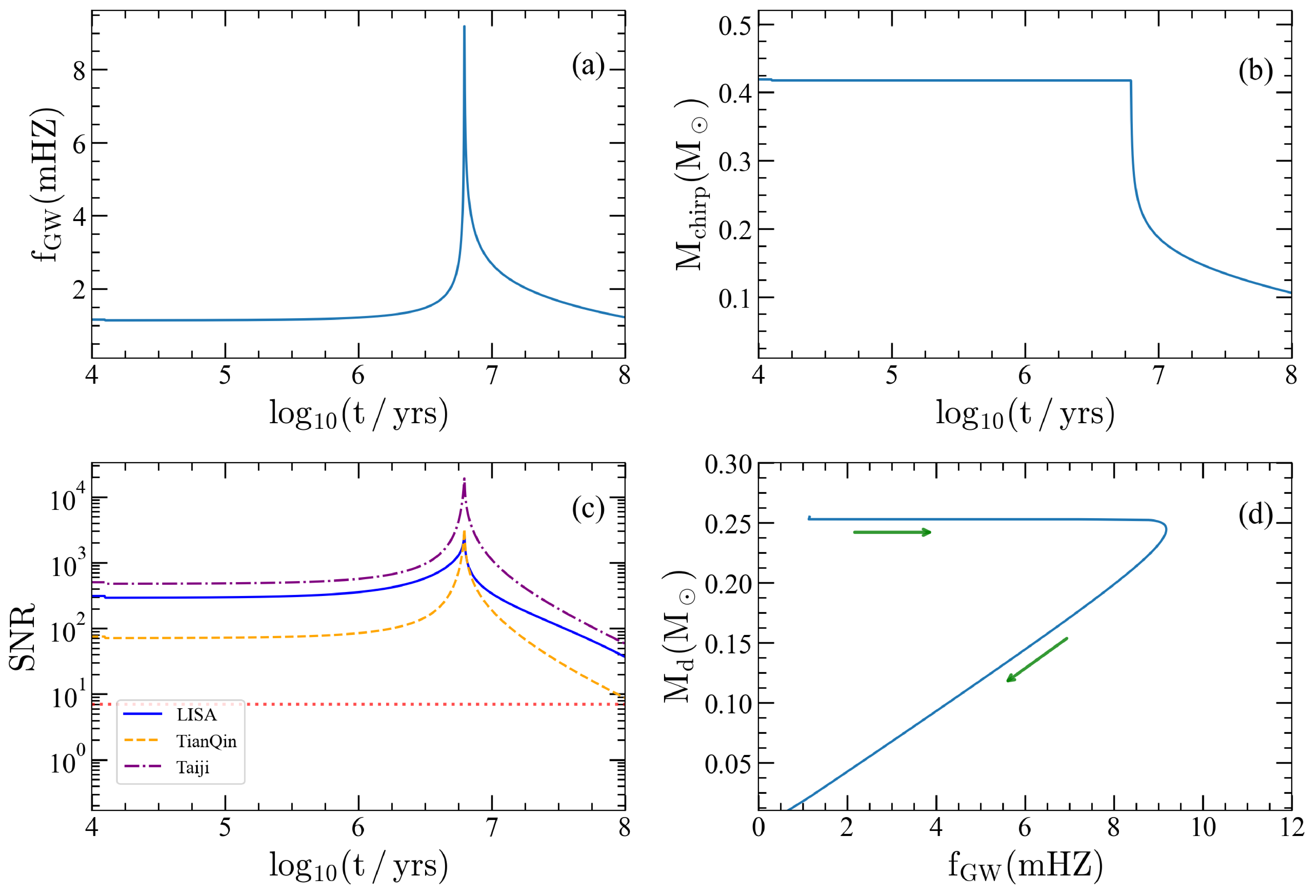}
    \caption{Temporal evolution of the GW frequency (\textit{panel~a}), chirp mass (\textit{panel~b}) and SNR (\textit{panel~c}) from our binary evolution calculation. The donor mass vs. GW frequency is also plotted (\textit{panel~d}). The horizontal dotted line indicates the critical value of $\rm{SNR=7}$, above which the source will be detectable for \textit{LISA}, \textit{TianQin} and \textit{Taiji}. The SNR is computed for a mission lifetime of 4 years, which is the nominal operational period for \textit{LISA}, \textit{TianQin} and \textit{Taiji}}.
    \label{fig:GW}
\end{figure}

\begin{figure}
    \centering
    \includegraphics[width=0.88\linewidth]{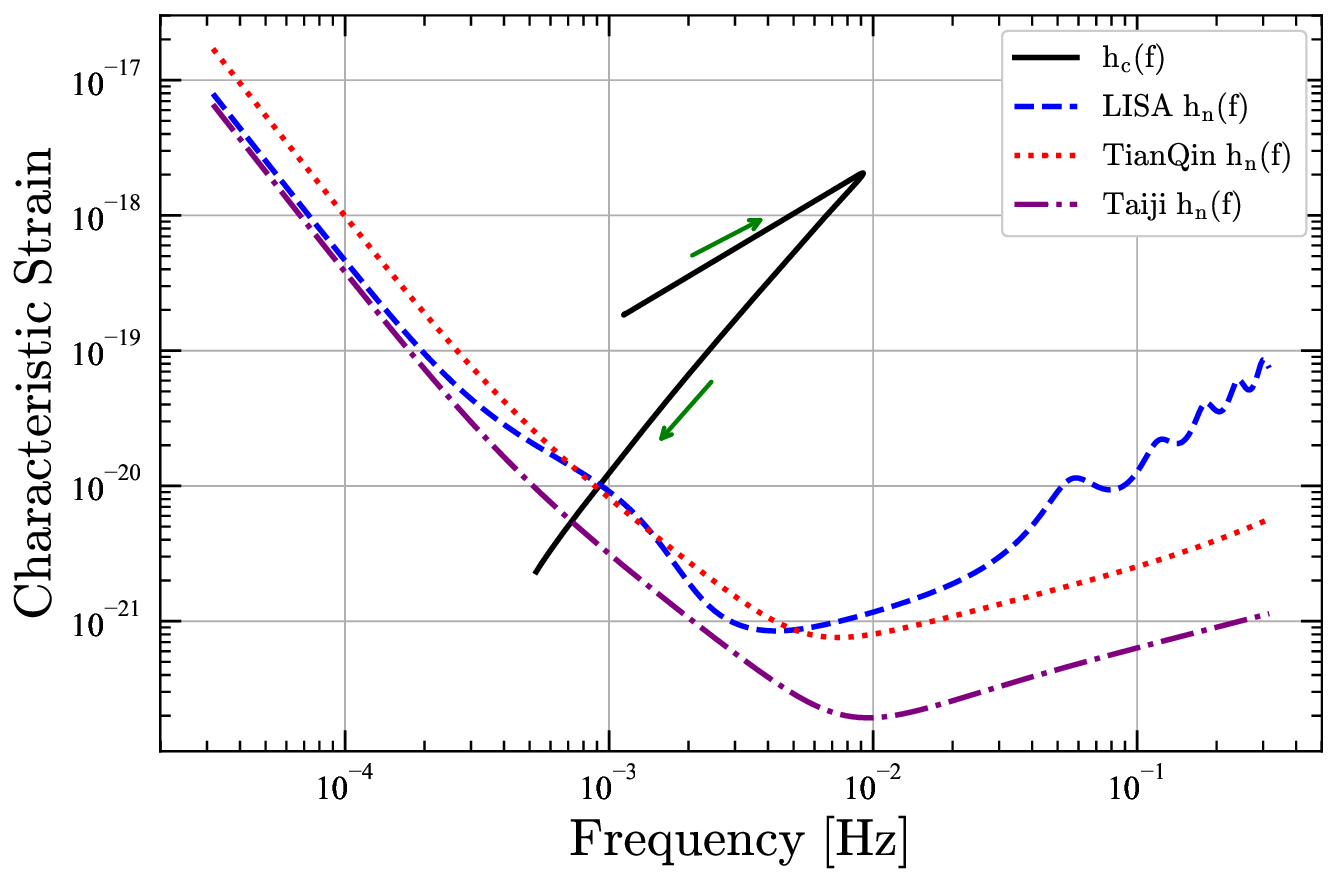}
    \caption{Characteristic strain of the GW signal expected from ATLAS~J1138-5139, which is compared to the sensitivity curves of \textit{LISA} (i.e., the red dotted line), \textit{TianQin} (i.e., the blue dashed line) and \textit{Taiji} (i.e., the purple dashdotted line). The GW signal curve of ATLAS~J1138-5139 (black solid line) lies above the sensitivity thresholds, suggesting that the GW signal of ATLAS~J1138-5139 would be likely to be detected by \textit{LISA}, \textit{TianQin} and \textit{Taiji}.}
    \label{fig:lisa}
\end{figure}

In Figure~\ref{fig:GW}, we present the time evolution of the GW frequency, chirp mass and signal-to-noise ratio (SNR), in which the evolution of donor mass as a function of GW frequency is also shown. Here the SNR for \textit{LISA}, \textit{TianQin} and \textit{Taiji} is computed with the Python package LEGWORK \citep{Wagg_2022}. It shows that the binary system has a strong GW emission in the mHz regime (see top-left panel of Fig.~\ref{fig:GW}), and the SNR can be up to $\sim 800$ (bottom-left panel of Fig.~\ref{fig:GW}) at a distance of $553\;{\rm pc}$ for ATLAS~J1138-5139. Given the detection threshold of $\rm{SNR=7}$ for \textit{LISA}, \textit{TianQin} and \textit{Taiji}, we expect that ATLAS~J1138-5139 will be detectable to the \textit{LISA}, \textit{TianQin} and \textit{Taiji}. In Figure~\ref{fig:lisa}, it further shows that the characteristic strain of this binary system lies significantly above the nominal sensitivity curves of \textit{LISA}, \textit{TianQin} and \textit{Taiji}, which indicates that the GW signal from ATLAS~J1138-5139 will be very likely to be detected by \textit{LISA}, \textit{TianQin} and \textit{Taiji}.

\subsection{Uncertainties of our calculations}

Our simulations show that the accumulated He-shell mass reaches about $0.12\,M_{\odot}$. This value falls within the range commonly used for triggering a DDet explosion of a WD for SNe~Ia ($0.05\,M_{\odot}-0.15\,M_{\odot}$; \citealt{Fink2007, Fink2010,Sim2010,2011ApJ...734...38W,Shen2018,Shen2021,Townsley2019,Boos2021, Gronow2021, Collins2022, Wong2023,Zenati_2023,Liu_2023}). For a CO~WD with a mass of $\sim 1.0\,M_{\odot}$, previous studies suggest that a He-shell mass of $\lesssim 0.1\,M_{\odot}$ might be sufficient to cause a successful DDet explosion \citep{Wang_2013,Wu_2017}. We therefore simply assume that our model would be likely to trigger a DDet explosion to cause a SN~Ia eventually once the He-shell mass reaches $\sim 0.1\,M_{\odot}$.

Following some recent studies \citep{2023JApA...44...35K,2024A&A...686A.227P,2025A&A...704A..82R,2025A&A...700A.107G}, we simply adopt the He accumulation efficiency given by \cite{Piersanti_2014}. However, the exact He accumulation efficiency is still poorly constrained. This might lead to some uncertainties of our results. For example, some previous studies have pointed out that the critical accretion rates separating different accumulation regimes may depend on the WD mass, composition, and thermal state \citep[e.g.,][]{Kato_2004,2016RAA....16..160W,2014MNRAS.440L.101R,2014arXiv1403.4797T,ruiter2024typeiasupernovaprogenitors}. However, as above mentioned, the accumulated He-shell masses onto the WD could reach to $0.12\,M_{\odot}$ in our calculations. This He-shell mass is larger than the typical critical He-shell for a DDet explosion \citep{Wang_2013,Wu_2017}. We therefore do not expect that uncertainties on the He accumulation efficiency will change the conclusions of this work. In this future study, we will investigate the effect of different He accumulation models on our results.


\section{Summary and Conclusions}
\label{sect:discussion}

In this study, we investigate the evolution, fate and GW emission characteristics of the newly detected DWD system ATLAS~J1138-5139 by performing detailed binary evolution calculations with the \texttt{MESA} code. The starting model of our binary evolution calculation is selected to be consistent with the observed binary properties of ATLAS~J1138-5139, i.e., it consists of a $0.255\,M_{\odot}$ He~WD (which contains a $0.015\,M_{\odot}$ H envelope) and a $0.99\,M_{\odot}$ CO~WD with an initial orbital period of $0.02\,\mathrm{days}$. Our main results and conclusions can be summarized as follows.

\begin{enumerate}
\item[(1)]
We find that ATLAS~J1138-5139 will evolve into an AM~CVn system in about $\sim6.3\,\mathrm{Myr}$. Afterwards, the transferred material from the He~WD companion can build up to form a He shell gradually near the surface of the CO~WD.

\item[(2)] 
Our detailed binary evolution calculations show that the accumulated He-shell masses onto the CO~WD can be up to approximately $0.12\,{M_{\odot}}$, indicating that the accreting CO~WD is likely to trigger a DDet explosion. We therefore conclude that ATLAS~J1138-5139 would be likely to eventually explode as an SN~Ia. 

\item[(3)] 
Our calculations suggest that ATLAS~J1138-5139 will be an important source for the GW detection by the upcoming space-based GW detectors like \textit{LISA}, \textit{TianQin} and \textit{Taiji} in its late-time evolution stages. The signal to noise ratio reaches about $800$ for a four-year mission lifetime of these GW detectors.

\end{enumerate}

\normalem
\begin{acknowledgements}
We are grateful to the anonymous referee for the valuable comments that helped us to improve this paper. This work has been supported by the National Natural Science Foundation of China (NSFC, Nos.\ 12288102, 12125303, 12090040/3, 12090040/1, 11873016, 12333008 and 12422305), the Strategic Priority Research Program of the Chinese Academy of Sciences (grant Nos. XDB1160303, XDB1160300, XDB1160000, XDB1160200, XDB1160201), the National Key R\&D Program of China (Nos.\ 2021YFA1600403, 2021YFA1600401 and 2021YFA1600400), the International Centre of Supernovae (ICESUN), Yunnan Key Laboratory of Supernova Research (No. 202505AV340004), the Yunnan Fundamental Research Projects (grant Nos.\ 202201BC070003, 202001AW070007), the Yunnan Revitalization Talent Support Program "YunLing Scholar" project, the New Cornerstone Science Foundation through the XPLORER PRlZE, the CAS Project for Young Scientists in Basic Research (YSBR-148), the Young Talent Project of Yunnan Revitalization Talent Support Program and the Yunnan Revitalization Talent Support Program-Science \& Technology Champion Project (No.202305AB350003). 
\end{acknowledgements} 
  
\bibliographystyle{raa}
\bibliography{bibtex}

@article{Brown_2011,
doi = {10.1088/2041-8205/737/1/L23},
url = {https://doi.org/10.1088/2041-8205/737/1/L23},
year = {2011},
month = {jul},
publisher = {The American Astronomical Society},
volume = {737},
number = {1},
pages = {L23},
author = {Brown, Warren R. and Kilic, Mukremin and Hermes, J. J. and Prieto, Carlos Allende and Kenyon, Scott J. and Winget, D. E.},
title = {A 12 MINUTE ORBITAL PERIOD DETACHED WHITE DWARF ECLIPSING BINARY*},
journal = {The Astrophysical Journal Letters}
}

@article{Brown_2020,
doi = {10.3847/2041-8213/ab8228},
url = {https://doi.org/10.3847/2041-8213/ab8228},
year = {2020},
month = {apr},
publisher = {The American Astronomical Society},
volume = {892},
number = {2},
pages = {L35},
author = {Brown, Warren R. and Kilic, Mukremin and Bédard, A. and Kosakowski, Alekzander and Bergeron, P.},
title = {A 1201 s Orbital Period Detached Binary: The First Double Helium Core White Dwarf LISA Verification Binary},
journal = {The Astrophysical Journal Letters}
}

@ARTICLE{1995ApJ...438..887R,
       author = {{Rasio}, Frederic A. and {Shapiro}, Stuart L.},
        title = "{Hydrodynamics of Binary Coalescence. II. Polytropes with Gamma = 5/3}",
      journal = {\apj},
         year = 1995,
        month = jan,
       volume = {438},
        pages = {887},
          doi = {10.1086/175130},
archivePrefix = {arXiv},
       eprint = {astro-ph/9406032},
 primaryClass = {astro-ph},
       adsurl = {https://ui.adsabs.harvard.edu/abs/1995ApJ...438..887R}
}

@ARTICLE{1997ApJ...481..355S,
       author = {{Segretain}, Laurent and {Chabrier}, Gilles and {Mochkovitch}, Robert},
        title = "{The Fate of Merging White Dwarfs}",
      journal = {\apj},
         year = 1997,
        month = may,
       volume = {481},
       number = {1},
        pages = {355-362},
          doi = {10.1086/304015},
       adsurl = {https://ui.adsabs.harvard.edu/abs/1997ApJ...481..355S}
}

@ARTICLE{2009A&A...500.1193L,
       author = {{Lor{\'e}n-Aguilar}, P. and {Isern}, J. and {Garc{\'\i}a-Berro}, E.},
        title = "{High-resolution smoothed particle hydrodynamics simulations of the merger of binary white dwarfs}",
      journal = {\aap},
         year = 2009,
        month = jun,
       volume = {500},
       number = {3},
        pages = {1193-1205},
          doi = {10.1051/0004-6361/200811060},
       adsurl = {https://ui.adsabs.harvard.edu/abs/2009A&A...500.1193L}
}

@ARTICLE{2010Natur.463...61P,
       author = {{Pakmor}, R{\"u}diger and {Kromer}, Markus and {R{\"o}pke}, Friedrich K. and {Sim}, Stuart A. and {Ruiter}, Ashley J. and {Hillebrandt}, Wolfgang},
        title = "{Sub-luminous type Ia supernovae from the mergers of equal-mass white dwarfs with mass \raisebox{-0.5ex}\textasciitilde0.9M$_{solar}$}",
      journal = {\nat},
         year = 2010,
        month = jan,
       volume = {463},
       number = {7277},
        pages = {61-64},
          doi = {10.1038/nature08642},
archivePrefix = {arXiv},
       eprint = {0911.0926},
 primaryClass = {astro-ph.HE},
       adsurl = {https://ui.adsabs.harvard.edu/abs/2010Natur.463...61P}
}

@ARTICLE{2015ApJ...807..105S,
       author = {{Sato}, Yushi and {Nakasato}, Naohito and {Tanikawa}, Ataru and {Nomoto}, Ken'ichi and {Maeda}, Keiichi and {Hachisu}, Izumi},
        title = "{A Systematic Study of Carbon-Oxygen White Dwarf Mergers: Mass Combinations for Type Ia Supernovae}",
      journal = {\apj},
         year = 2015,
        month = jul,
       volume = {807},
       number = {1},
          eid = {105},
        pages = {105},
          doi = {10.1088/0004-637X/807/1/105},
archivePrefix = {arXiv},
       eprint = {1505.01646},
 primaryClass = {astro-ph.HE},
       adsurl = {https://ui.adsabs.harvard.edu/abs/2015ApJ...807..105S}
}

@ARTICLE{2012ApJ...747L..10P,
       author = {{Pakmor}, R. and {Kromer}, M. and {Taubenberger}, S. and {Sim}, S.~A. and {R{\"o}pke}, F.~K. and {Hillebrandt}, W.},
        title = "{Normal Type Ia Supernovae from Violent Mergers of White Dwarf Binaries}",
      journal = {\apjl},
         year = 2012,
        month = mar,
       volume = {747},
       number = {1},
          eid = {L10},
        pages = {L10},
          doi = {10.1088/2041-8205/747/1/L10},
archivePrefix = {arXiv},
       eprint = {1201.5123},
 primaryClass = {astro-ph.HE},
       adsurl = {https://ui.adsabs.harvard.edu/abs/2012ApJ...747L..10P}
}

@ARTICLE{2010ApJ...725..296F,
       author = {{Fryer}, Chris L. and {Ruiter}, Ashley J. and {Belczynski}, Krzysztof and {Brown}, Peter J. and {Bufano}, Filomena and {Diehl}, Steven and {Fontes}, Christopher J. and {Frey}, Lucille H. and {Holland}, Stephen T. and {Hungerford}, Aimee L. and {Immler}, Stefan and {Mazzali}, Paolo and {Meakin}, Casey and {Milne}, Peter A. and {Raskin}, Cody and {Timmes}, Francis X.},
        title = "{Spectra of Type Ia Supernovae from Double Degenerate Mergers}",
      journal = {\apj},
         year = 2010,
        month = dec,
       volume = {725},
       number = {1},
        pages = {296-308},
          doi = {10.1088/0004-637X/725/1/296},
archivePrefix = {arXiv},
       eprint = {1007.0570},
 primaryClass = {astro-ph.SR},
       adsurl = {https://ui.adsabs.harvard.edu/abs/2010ApJ...725..296F}
}

@ARTICLE{2004A&A...413..257G,
       author = {{Guerrero}, J. and {Garc{\'\i}a-Berro}, E. and {Isern}, J.},
        title = "{Smoothed Particle Hydrodynamics simulations  of merging white dwarfs}",
      journal = {\aap},
         year = 2004,
        month = jan,
       volume = {413},
        pages = {257-272},
          doi = {10.1051/0004-6361:20031504},
       adsurl = {https://ui.adsabs.harvard.edu/abs/2004A&A...413..257G}
}

@article{Brown_2022,
doi = {10.3847/1538-4357/ac72ac},
url = {https://doi.org/10.3847/1538-4357/ac72ac},
year = {2022},
month = {jul},
publisher = {The American Astronomical Society},
volume = {933},
number = {1},
pages = {94},
author = {Brown, Warren R. and Kilic, Mukremin and Kosakowski, Alekzander and Gianninas, A.},
title = {The ELM Survey. IX. A Complete Sample of Low-mass White Dwarf Binaries in the SDSS Footprint},
journal = {The Astrophysical Journal}
}

@article{Kilic_2021,
doi = {10.3847/2041-8213/ac1e2b},
url = {https://doi.org/10.3847/2041-8213/ac1e2b},
year = {2021},
month = {aug},
publisher = {The American Astronomical Society},
volume = {918},
number = {1},
pages = {L14},
author = {Kilic, Mukremin and Brown, Warren R. and Bédard, A. and Kosakowski, Alekzander},
title = {The Discovery of Two LISA Sources within 0.5 kpc},
journal = {The Astrophysical Journal Letters}
}

@article{Bellm_2019,
doi = {10.1088/1538-3873/aaecbe},
url = {https://doi.org/10.1088/1538-3873/aaecbe},
year = {2018},
month = {dec},
publisher = {The Astronomical Society of the Pacific},
volume = {131},
number = {995},
pages = {018002},
author = {Bellm, Eric C. and Kulkarni, Shrinivas R. and Graham, Matthew J. and Dekany, Richard and Smith, Roger M. and Riddle, Reed and Masci, Frank J. and Helou, George and Prince, Thomas A. and Adams, Scott M. and Barbarino, C. and Barlow, Tom and Bauer, James and Beck, Ron and Belicki, Justin and Biswas, Rahul and Blagorodnova, Nadejda and Bodewits, Dennis and Bolin, Bryce and Brinnel, Valery and Brooke, Tim and Bue, Brian and Bulla, Mattia and Burruss, Rick and Cenko, S. Bradley and Chang, Chan-Kao and Connolly, Andrew and Coughlin, Michael and Cromer, John and Cunningham, Virginia and De, Kishalay and Delacroix, Alex and Desai, Vandana and Duev, Dmitry A. and Eadie, Gwendolyn and Farnham, Tony L. and Feeney, Michael and Feindt, Ulrich and Flynn, David and Franckowiak, Anna and Frederick, S. and Fremling, C. and Gal-Yam, Avishay and Gezari, Suvi and Giomi, Matteo and Goldstein, Daniel A. and Golkhou, V. Zach and Goobar, Ariel and Groom, Steven and Hacopians, Eugean and Hale, David and Henning, John and Ho, Anna Y. Q. and Hover, David and Howell, Justin and Hung, Tiara and Huppenkothen, Daniela and Imel, David and Ip, Wing-Huen and Ivezić, Željko and Jackson, Edward and Jones, Lynne and Juric, Mario and Kasliwal, Mansi M. and Kaspi, S. and Kaye, Stephen and Kelley, Michael S. P. and Kowalski, Marek and Kramer, Emily and Kupfer, Thomas and Landry, Walter and Laher, Russ R. and Lee, Chien-De and Lin, Hsing Wen and Lin, Zhong-Yi and Lunnan, Ragnhild and Giomi, Matteo and Mahabal, Ashish and Mao, Peter and Miller, Adam A. and Monkewitz, Serge and Murphy, Patrick and Ngeow, Chow-Choong and Nordin, Jakob and Nugent, Peter and Ofek, Eran and Patterson, Maria T. and Penprase, Bryan and Porter, Michael and Rauch, Ludwig and Rebbapragada, Umaa and Reiley, Dan and Rigault, Mickael and Rodriguez, Hector and Roestel, Jan van and Rusholme, Ben and Santen, Jakob van and Schulze, S. and Shupe, David L. and Singer, Leo P. and Soumagnac, Maayane T. and Stein, Robert and Surace, Jason and Sollerman, Jesper and Szkody, Paula and Taddia, F. and Terek, Scott and Van Sistine, Angela and van Velzen, Sjoert and Vestrand, W. Thomas and Walters, Richard and Ward, Charlotte and Ye, Quan-Zhi and Yu, Po-Chieh and Yan, Lin and Zolkower, Jeffry},
title = {The Zwicky Transient Facility: System Overview, Performance, and First Results},
journal = {Publications of the Astronomical Society of the Pacific}
}

@ARTICLE{1990ApJ...348..647B,
author = {{Benz}, W. and {Bowers}, R.~L. and {Cameron}, A.~G.~W. and {Press}, W.~H. .},
title = "{Dynamic Mass Exchange in Doubly Degenerate Binaries. I. 0.9 and 1.2 M$_{sun}$ Stars}",
journal = {\apj},
year = 1990,
month = jan,
volume = {348},
pages = {647},
doi = {10.1086/168273},
adsurl = {https://ui.adsabs.harvard.edu/abs/1990ApJ...348..647B}
}

@article{Graham_2019,
doi = {10.1088/1538-3873/ab006c},
url = {https://doi.org/10.1088/1538-3873/ab006c},
year = {2019},
month = {may},
publisher = {The Astronomical Society of the Pacific},
volume = {131},
number = {1001},
pages = {078001},
author = {Graham, Matthew J. and Kulkarni, S. R. and Bellm, Eric C. and Adams, Scott M. and Barbarino, Cristina and Blagorodnova, Nadejda and Bodewits, Dennis and Bolin, Bryce and Brady, Patrick R. and Cenko, S. Bradley and Chang, Chan-Kao and Coughlin, Michael W. and De, Kishalay and Eadie, Gwendolyn and Farnham, Tony L. and Feindt, Ulrich and Franckowiak, Anna and Fremling, Christoffer and Gezari, Suvi and Ghosh, Shaon and Goldstein, Daniel A. and Golkhou, V. Zach and Goobar, Ariel and Ho, Anna Y. Q. and Huppenkothen, Daniela and Ivezić, Željko and Jones, R. Lynne and Juric, Mario and Kaplan, David L. and Kasliwal, Mansi M. and Kelley, Michael S. P. and Kupfer, Thomas and Lee, Chien-De and Lin, Hsing Wen and Lunnan, Ragnhild and Mahabal, Ashish A. and Miller, Adam A. and Ngeow, Chow-Choong and Nugent, Peter and Ofek, Eran O. and Prince, Thomas A. and Rauch, Ludwig and Roestel, Jan van and Schulze, Steve and Singer, Leo P. and Sollerman, Jesper and Taddia, Francesco and Yan, Lin and Ye, Quan-Zhi and Yu, Po-Chieh and Barlow, Tom and Bauer, James and Beck, Ron and Belicki, Justin and Biswas, Rahul and Brinnel, Valery and Brooke, Tim and Bue, Brian and Bulla, Mattia and Burruss, Rick and Connolly, Andrew and Cromer, John and Cunningham, Virginia and Dekany, Richard and Delacroix, Alex and Desai, Vandana and Duev, Dmitry A. and Feeney, Michael and Flynn, David and Frederick, Sara and Gal-Yam, Avishay and Giomi, Matteo and Groom, Steven and Hacopians, Eugean and Hale, David and Helou, George and Henning, John and Hover, David and Hillenbrand, Lynne A. and Howell, Justin and Hung, Tiara and Imel, David and Ip, Wing-Huen and Jackson, Edward and Kaspi, Shai and Kaye, Stephen and Kowalski, Marek and Kramer, Emily and Kuhn, Michael and Landry, Walter and Laher, Russ R. and Mao, Peter and Masci, Frank J. and Monkewitz, Serge and Murphy, Patrick and Nordin, Jakob and Patterson, Maria T. and Penprase, Bryan and Porter, Michael and Rebbapragada, Umaa and Reiley, Dan and Riddle, Reed and Rigault, Mickael and Rodriguez, Hector and Rusholme, Ben and Santen, Jakob van and Shupe, David L. and Smith, Roger M. and Soumagnac, Maayane T. and Stein, Robert and Surace, Jason and Szkody, Paula and Terek, Scott and Sistine, Angela Van and Velzen, Sjoert van and Vestrand, W. Thomas and Walters, Richard and Ward, Charlotte and Zhang, Chaoran and Zolkower, Jeffry},
title = {The Zwicky Transient Facility: Science Objectives},
journal = {Publications of the Astronomical Society of the Pacific}
}

@article{Masci_2019,
doi = {10.1088/1538-3873/aae8ac},
url = {https://doi.org/10.1088/1538-3873/aae8ac},
year = {2018},
month = {dec},
publisher = {The Astronomical Society of the Pacific},
volume = {131},
number = {995},
pages = {018003},
author = {Masci, Frank J. and Laher, Russ R. and Rusholme, Ben and Shupe, David L. and Groom, Steven and Surace, Jason and Jackson, Edward and Monkewitz, Serge and Beck, Ron and Flynn, David and Terek, Scott and Landry, Walter and Hacopians, Eugean and Desai, Vandana and Howell, Justin and Brooke, Tim and Imel, David and Wachter, Stefanie and Ye, Quan-Zhi and Lin, Hsing-Wen and Cenko, S. Bradley and Cunningham, Virginia and Rebbapragada, Umaa and Bue, Brian and Miller, Adam A. and Mahabal, Ashish and Bellm, Eric C. and Patterson, Maria T. and Jurić, Mario and Golkhou, V. Zach and Ofek, Eran O. and Walters, Richard and Graham, Matthew and Kasliwal, Mansi M. and Dekany, Richard G. and Kupfer, Thomas and Burdge, Kevin and Cannella, Christopher B. and Barlow, Tom and Sistine, Angela Van and Giomi, Matteo and Fremling, Christoffer and Blagorodnova, Nadejda and Levitan, David and Riddle, Reed and Smith, Roger M. and Helou, George and Prince, Thomas A. and Kulkarni, Shrinivas R.},
title = {The Zwicky Transient Facility: Data Processing, Products, and Archive},
journal = {Publications of the Astronomical Society of the Pacific}
}

@article{Kilic_2014,
   title={A new gravitational wave verification source},
   volume={444},
   ISSN={1745-3925},
   url={http://dx.doi.org/10.1093/mnrasl/slu093},
   DOI={10.1093/mnrasl/slu093},
   number={1},
   journal={Monthly Notices of the Royal Astronomical Society: Letters},
   publisher={Oxford University Press (OUP)},
   author={Kilic, Mukremin and Brown, Warren R. and Gianninas, A. and Hermes, J. J. and Allende Prieto, Carlos and Kenyon, S. J.},
   year={2014},
   month=jul, pages={L1–L5} }

@article{Wagg_2022,
   title={LEGWORK: A Python Package for Computing the Evolution and Detectability of Stellar-origin Gravitational-wave Sources with Space-based Detectors},
   volume={260},
   ISSN={1538-4365},
   url={http://dx.doi.org/10.3847/1538-4365/ac5c52},
   DOI={10.3847/1538-4365/ac5c52},
   number={2},
   journal={The Astrophysical Journal Supplement Series},
   publisher={American Astronomical Society},
   author={Wagg, T. and Breivik, K. and de Mink, S. E.},
   year={2022},
   month=jun, pages={52} }

@INPROCEEDINGS{2014IAUS..298..269D,
       author = {{Deng}, Licai},
        title = "{LAMOST Experiment on Galactic Understanding and Exploration: An overview}",
    booktitle = {Setting the scene for Gaia and LAMOST},
         year = 2014,
       editor = {{Feltzing}, Sofia and {Zhao}, Gang and {Walton}, Nicholas A. and {Whitelock}, Patricia},
       series = {IAU Symposium},
       volume = {298},
        month = jan,
        pages = {269-280},
          doi = {10.1017/S1743921313006467},
       adsurl = {https://ui.adsabs.harvard.edu/abs/2014IAUS..298..269D}
}

@misc{ebadi2024lisadoublewhitedwarf,
      title={LISA double white dwarf binaries as Galactic accelerometers}, 
      author={Reza Ebadi and Vladimir Strokov and Erwin H. Tanin and Emanuele Berti and Ronald L. Walsworth},
      year={2024},
      eprint={2405.13109},
      archivePrefix={arXiv},
      primaryClass={gr-qc},
      url={https://arxiv.org/abs/2405.13109}, 
}

@article{Liu_2022,
   title={Signatures of a Surviving Helium-star Companion in Type Ia Supernovae and Constraints on the Progenitor Companion of SN 2011fe},
   volume={928},
   ISSN={1538-4357},
   url={http://dx.doi.org/10.3847/1538-4357/ac5517},
   DOI={10.3847/1538-4357/ac5517},
   number={2},
   journal={The Astrophysical Journal},
   publisher={American Astronomical Society},
   author={Liu, Zheng-Wei and Röpke, Friedrich K. and Zeng, Yaotian},
   year={2022},
   month=apr, pages={146} }

@article{Burdge_2020,
   title={An 8.8 Minute Orbital Period Eclipsing Detached Double White Dwarf Binary},
   volume={905},
   ISSN={2041-8213},
   url={http://dx.doi.org/10.3847/2041-8213/abca91},
   DOI={10.3847/2041-8213/abca91},
   number={1},
   journal={The Astrophysical Journal Letters},
   publisher={American Astronomical Society},
   author={Burdge, Kevin B. and Coughlin, Michael W. and Fuller, Jim and Kaplan, David L. and Kulkarni, S. R. and Marsh, Thomas R. and Bellm, Eric C. and Dekany, Richard G. and Duev, Dmitry A. and Graham, Matthew J. and Mahabal, Ashish A. and Masci, Frank J. and Laher, Russ R. and Riddle, Reed and Soumagnac, Maayane T. and Prince, Thomas A.},
   year={2020},
   month=dec, pages={L7} }

@misc{amaroseoane2017laserinterferometerspaceantenna,
      title={Laser Interferometer Space Antenna}, 
      author={Pau Amaro-Seoane and Heather Audley and Stanislav Babak and John Baker and Enrico Barausse and Peter Bender and Emanuele Berti and Pierre Binetruy and Michael Born and Daniele Bortoluzzi and Jordan Camp and Chiara Caprini and Vitor Cardoso and Monica Colpi and John Conklin and Neil Cornish and Curt Cutler and Karsten Danzmann and Rita Dolesi and Luigi Ferraioli and Valerio Ferroni and Ewan Fitzsimons and Jonathan Gair and Lluis Gesa Bote and Domenico Giardini and Ferran Gibert and Catia Grimani and Hubert Halloin and Gerhard Heinzel and Thomas Hertog and Martin Hewitson and Kelly Holley-Bockelmann and Daniel Hollington and Mauro Hueller and Henri Inchauspe and Philippe Jetzer and Nikos Karnesis and Christian Killow and Antoine Klein and Bill Klipstein and Natalia Korsakova and Shane L Larson and Jeffrey Livas and Ivan Lloro and Nary Man and Davor Mance and Joseph Martino and Ignacio Mateos and Kirk McKenzie and Sean T McWilliams and Cole Miller and Guido Mueller and Germano Nardini and Gijs Nelemans and Miquel Nofrarias and Antoine Petiteau and Paolo Pivato and Eric Plagnol and Ed Porter and Jens Reiche and David Robertson and Norna Robertson and Elena Rossi and Giuliana Russano and Bernard Schutz and Alberto Sesana and David Shoemaker and Jacob Slutsky and Carlos F. Sopuerta and Tim Sumner and Nicola Tamanini and Ira Thorpe and Michael Troebs and Michele Vallisneri and Alberto Vecchio and Daniele Vetrugno and Stefano Vitale and Marta Volonteri and Gudrun Wanner and Harry Ward and Peter Wass and William Weber and John Ziemer and Peter Zweifel},
      year={2017},
      eprint={1702.00786},
      archivePrefix={arXiv},
      primaryClass={astro-ph.IM},
      url={https://arxiv.org/abs/1702.00786}, 
}

@article{Luo_2016,
   title={TianQin: a space-borne gravitational wave detector},
   volume={33},
   ISSN={1361-6382},
   url={http://dx.doi.org/10.1088/0264-9381/33/3/035010},
   DOI={10.1088/0264-9381/33/3/035010},
   number={3},
   journal={Classical and Quantum Gravity},
   publisher={IOP Publishing},
   author={Luo, Jun and Chen, Li-Sheng and Duan, Hui-Zong and Gong, Yun-Gui and Hu, Shoucun and Ji, Jianghui and Liu, Qi and Mei, Jianwei and Milyukov, Vadim and Sazhin, Mikhail and Shao, Cheng-Gang and Toth, Viktor T and Tu, Hai-Bo and Wang, Yamin and Wang, Yan and Yeh, Hsien-Chi and Zhan, Ming-Sheng and Zhang, Yonghe and Zharov, Vladimir and Zhou, Ze-Bing},
   year={2016},
   month=jan, pages={035010} }

@article{Ruan_2020,
   title={Taiji program: Gravitational-wave sources},
   volume={35},
   ISSN={1793-656X},
   url={http://dx.doi.org/10.1142/S0217751X2050075X},
   DOI={10.1142/s0217751x2050075x},
   number={17},
   journal={International Journal of Modern Physics A},
   publisher={World Scientific Pub Co Pte Lt},
   author={Ruan, Wen-Hong and Guo, Zong-Kuan and Cai, Rong-Gen and Zhang, Yuan-Zhong},
   year={2020},
   month=jun, pages={2050075} }

@ARTICLE{2010PASP..122.1133S,
       author = {{Solheim}, J. -E.},
        title = "{AM CVn Stars: Status and Challenges}",
      journal = {\pasp},
         year = 2010,
        month = oct,
       volume = {122},
       number = {896},
        pages = {1133},
          doi = {10.1086/656680},
       adsurl = {https://ui.adsabs.harvard.edu/abs/2010PASP..122.1133S}
}

@article{Nelemans_2004,
   title={Short-period AM CVn systems as optical, X-ray and gravitational-wave sources},
   volume={349},
   ISSN={0035-8711},
   url={http://dx.doi.org/10.1111/j.1365-2966.2004.07479.x},
   DOI={10.1111/j.1365-2966.2004.07479.x},
   number={1},
   journal={Monthly Notices of the Royal Astronomical Society},
   publisher={Oxford University Press (OUP)},
   author={Nelemans, G. and Yungelson, L. R. and Portegies Zwart, S. F.},
   year={2004},
   month=mar, pages={181–192} }

@ARTICLE{2002ARep...46..667T,
       author = {{Tutukov}, A.~V. and {Yungelson}, L.~R.},
        title = "{A Model for the Population of Binary Stars in the Galaxy}",
      journal = {Astronomy Reports},
         year = 2002,
        month = aug,
       volume = {46},
       number = {8},
        pages = {667-683},
          doi = {10.1134/1.1502227},
       adsurl = {https://ui.adsabs.harvard.edu/abs/2002ARep...46..667T}
}

@ARTICLE{Wong2023,
       author = {{Wong}, Tin Long Sunny and {Bildsten}, Lars},
        title = "{Dynamical He Flashes in Double White Dwarf Binaries}",
      journal = {\apj},
         year = 2023,
        month = jul,
       volume = {951},
       number = {1},
          eid = {28},
        pages = {28},
          doi = {10.3847/1538-4357/acce9d},
archivePrefix = {arXiv},
       eprint = {2305.05695},
 primaryClass = {astro-ph.SR},
       adsurl = {https://ui.adsabs.harvard.edu/abs/2023ApJ...951...28W}
}

@ARTICLE{Shen2021,
       author = {{Shen}, Ken J. and {Boos}, Samuel J. and {Townsley}, Dean M. and {Kasen}, Daniel},
        title = "{Multidimensional Radiative Transfer Calculations of Double Detonations of Sub-Chandrasekhar-mass White Dwarfs}",
      journal = {\apj},
         year = 2021,
        month = nov,
       volume = {922},
       number = {1},
          eid = {68},
        pages = {68},
          doi = {10.3847/1538-4357/ac2304},
archivePrefix = {arXiv},
       eprint = {2108.12435},
 primaryClass = {astro-ph.SR},
       adsurl = {https://ui.adsabs.harvard.edu/abs/2021ApJ...922...68S}
}

@ARTICLE{Townsley2019,
       author = {{Townsley}, Dean M. and {Miles}, Broxton J. and {Shen}, Ken J. and {Kasen}, Daniel},
        title = "{Double Detonations with Thin, Modestly Enriched Helium Layers can Make Normal Type Ia Supernovae}",
      journal = {\apjl},
         year = 2019,
        month = jun,
       volume = {878},
       number = {2},
          eid = {L38},
        pages = {L38},
          doi = {10.3847/2041-8213/ab27cd},
archivePrefix = {arXiv},
       eprint = {1903.10960},
 primaryClass = {astro-ph.SR},
       adsurl = {https://ui.adsabs.harvard.edu/abs/2019ApJ...878L..38T}
}

@ARTICLE{Shen2018,
       author = {{Shen}, Ken J. and {Kasen}, Daniel and {Miles}, Broxton J. and {Townsley}, Dean M.},
        title = "{Sub-Chandrasekhar-mass White Dwarf Detonations Revisited}",
      journal = {\apj},
         year = 2018,
        month = feb,
       volume = {854},
       number = {1},
          eid = {52},
        pages = {52},
          doi = {10.3847/1538-4357/aaa8de},
archivePrefix = {arXiv},
       eprint = {1706.01898},
 primaryClass = {astro-ph.HE},
       adsurl = {https://ui.adsabs.harvard.edu/abs/2018ApJ...854...52S}
}

@ARTICLE{Collins2022,
       author = {{Collins}, Christine E. and {Gronow}, Sabrina and {Sim}, Stuart A. and {R{\"o}pke}, Friedrich K.},
        title = "{Double detonations: variations in Type Ia supernovae due to different core and He shell masses - II. Synthetic observables}",
      journal = {\mnras},
         year = 2022,
        month = dec,
       volume = {517},
       number = {4},
        pages = {5289-5302},
          doi = {10.1093/mnras/stac2665},
archivePrefix = {arXiv},
       eprint = {2209.04305},
 primaryClass = {astro-ph.SR},
       adsurl = {https://ui.adsabs.harvard.edu/abs/2022MNRAS.517.5289C}
}

@ARTICLE{Boos2021,
       author = {{Boos}, Samuel J. and {Townsley}, Dean M. and {Shen}, Ken J. and {Caldwell}, Spencer and {Miles}, Broxton J.},
        title = "{Multidimensional Parameter Study of Double Detonation Type Ia Supernovae Originating from Thin Helium Shell White Dwarfs}",
      journal = {\apj},
         year = 2021,
        month = oct,
       volume = {919},
       number = {2},
          eid = {126},
        pages = {126},
          doi = {10.3847/1538-4357/ac07a2},
archivePrefix = {arXiv},
       eprint = {2101.12330},
 primaryClass = {astro-ph.HE},
       adsurl = {https://ui.adsabs.harvard.edu/abs/2021ApJ...919..126B}
}

@ARTICLE{Gronow2021,
       author = {{Gronow}, Sabrina and {Collins}, Christine E. and {Sim}, Stuart A. and {R{\"o}pke}, Friedrich K.},
        title = "{Double detonations of sub-M$_{Ch}$ CO white dwarfs: variations in Type Ia supernovae due to different core and He shell masses}",
      journal = {\aap},
         year = 2021,
        month = may,
       volume = {649},
          eid = {A155},
        pages = {A155},
          doi = {10.1051/0004-6361/202039954},
archivePrefix = {arXiv},
       eprint = {2102.06719},
 primaryClass = {astro-ph.SR},
       adsurl = {https://ui.adsabs.harvard.edu/abs/2021A&A...649A.155G}
}

@ARTICLE{Sim2010,
       author = {{Sim}, S.~A. and {R{\"o}pke}, F.~K. and {Hillebrandt}, W. and {Kromer}, M. and {Pakmor}, R. and {Fink}, M. and {Ruiter}, A.~J. and {Seitenzahl}, I.~R.},
        title = "{Detonations in Sub-Chandrasekhar-mass C+O White Dwarfs}",
      journal = {\apjl},
         year = 2010,
        month = may,
       volume = {714},
       number = {1},
        pages = {L52-L57},
          doi = {10.1088/2041-8205/714/1/L52},
archivePrefix = {arXiv},
       eprint = {1003.2917},
 primaryClass = {astro-ph.HE},
       adsurl = {https://ui.adsabs.harvard.edu/abs/2010ApJ...714L..52S}
}

@ARTICLE{Fink2010,
       author = {{Fink}, M. and {R{\"o}pke}, F.~K. and {Hillebrandt}, W. and {Seitenzahl}, I.~R. and {Sim}, S.~A. and {Kromer}, M.},
        title = "{Double-detonation sub-Chandrasekhar supernovae: can minimum helium shell masses detonate the core?}",
      journal = {\aap},
         year = 2010,
        month = may,
       volume = {514},
          eid = {A53},
        pages = {A53},
          doi = {10.1051/0004-6361/200913892},
archivePrefix = {arXiv},
       eprint = {1002.2173},
 primaryClass = {astro-ph.SR},
       adsurl = {https://ui.adsabs.harvard.edu/abs/2010A&A...514A..53F}
}

@ARTICLE{Fink2007,
       author = {{Fink}, M. and {Hillebrandt}, W. and {R{\"o}pke}, F.~K.},
        title = "{Double-detonation supernovae of sub-Chandrasekhar mass white dwarfs}",
      journal = {\aap},
         year = 2007,
        month = dec,
       volume = {476},
       number = {3},
        pages = {1133-1143},
          doi = {10.1051/0004-6361:20078438},
archivePrefix = {arXiv},
       eprint = {0710.5486},
 primaryClass = {astro-ph},
       adsurl = {https://ui.adsabs.harvard.edu/abs/2007A&A...476.1133F}
}

@article{Liu_2023,
   title={Type Ia Supernova Explosions in Binary Systems: A Review},
   volume={23},
   ISSN={1674-4527},
   url={http://dx.doi.org/10.1088/1674-4527/acd89e},
   DOI={10.1088/1674-4527/acd89e},
   number={8},
   journal={Research in Astronomy and Astrophysics},
   publisher={IOP Publishing},
   author={Liu, Zheng-Wei and Röpke, Friedrich K. and Han, Zhanwen},
   year={2023},
   month=jul, pages={082001} }

@misc{ruiter2024typeiasupernovaprogenitors,
      title={Type Ia supernova progenitors: a contemporary view of a long-standing puzzle}, 
      author={Ashley J. Ruiter and Ivo R. Seitenzahl},
      year={2024},
      eprint={2412.01766},
      archivePrefix={arXiv},
      primaryClass={astro-ph.SR},
      url={https://arxiv.org/abs/2412.01766}, 
}

@article{Chen_2020,
   title={Compact Intermediate-mass Black Hole X-Ray Binaries: Potential LISA Sources?},
   volume={896},
   ISSN={1538-4357},
   url={http://dx.doi.org/10.3847/1538-4357/ab9017},
   DOI={10.3847/1538-4357/ab9017},
   number={2},
   journal={The Astrophysical Journal},
   publisher={American Astronomical Society},
   author={Chen, Wen-Cong},
   year={2020},
   month=jun, pages={129} }

@ARTICLE{1980SSRv...27..563N,
       author = {{Nomoto}, K.},
        title = "{White dwarf models for type I supernovae and quiet supernovae, and presupernova evolution}",
      journal = {\ssr},
         year = 1980,
        month = nov,
       volume = {27},
       number = {3-4},
        pages = {563-570},
          doi = {10.1007/BF00168350},
       adsurl = {https://ui.adsabs.harvard.edu/abs/1980SSRv...27..563N}
}

@ARTICLE{2008AstL...34..620Y,
       author = {{Yungelson}, L.~R.},
        title = "{Evolution of low-mass helium stars in semidetached binaries}",
      journal = {Astronomy Letters},
         year = 2008,
        month = sep,
       volume = {34},
       number = {9},
        pages = {620-634},
          doi = {10.1134/S1063773708090053},
archivePrefix = {arXiv},
       eprint = {0804.2780},
 primaryClass = {astro-ph},
       adsurl = {https://ui.adsabs.harvard.edu/abs/2008AstL...34..620Y}
}

@ARTICLE{2021ApJ...922..245B,
       author = {{Bauer}, Evan B. and {Kupfer}, Thomas},
        title = "{Phases of Mass Transfer from Hot Subdwarfs to White Dwarf Companions and Their Photometric Properties}",
      journal = {\apj},
         year = 2021,
        month = dec,
       volume = {922},
       number = {2},
          eid = {245},
        pages = {245},
          doi = {10.3847/1538-4357/ac25f0},
archivePrefix = {arXiv},
       eprint = {2106.13297},
 primaryClass = {astro-ph.SR},
       adsurl = {https://ui.adsabs.harvard.edu/abs/2021ApJ...922..245B}
}

@ARTICLE{1994ApJ...423..371W,
       author = {{Woosley}, S.~E. and {Weaver}, Thomas A.},
        title = "{Sub--Chandrasekhar Mass Models for Type IA Supernovae}",
      journal = {\apj},
         year = 1994,
        month = mar,
       volume = {423},
        pages = {371},
          doi = {10.1086/173813},
       adsurl = {https://ui.adsabs.harvard.edu/abs/1994ApJ...423..371W}
}

@article{Guillochon_2010,
   title={SURFACE DETONATIONS IN DOUBLE DEGENERATE BINARY SYSTEMS TRIGGERED BY ACCRETION STREAM INSTABILITIES},
   volume={709},
   ISSN={2041-8213},
   url={http://dx.doi.org/10.1088/2041-8205/709/1/L64},
   DOI={10.1088/2041-8205/709/1/l64},
   number={1},
   journal={The Astrophysical Journal},
   publisher={American Astronomical Society},
   author={Guillochon, James and Dan, Marius and Ramirez-Ruiz, Enrico and Rosswog, Stephan},
   year={2010},
   month=jan, pages={L64–L69} }

\end{document}